\documentclass[twocolumn,showpacs,preprintnumbers,amsmath,amssymb,floatfix]{revtex4}

\usepackage{graphicx}
\usepackage{dcolumn}
\usepackage{bm}
\usepackage{float}

\begin{document}


\title{Self-similar Scale-free Networks and Disassortativity\\}

\author{Soon-Hyung Yook, Filippo Radicchi and Hildegard Meyer-Ortmanns}

 \email{h.ortmanns@iu-bremen.de}
\affiliation{%
SES, International University Bremen, P.O.Box 750561, D-28725 Bremen\\
}%

\date{\today}

\begin{abstract}
Self-similar networks with scale-free degree distribution have recently
attracted much attention, since these apparently incompatible properties were
reconciled in \cite{havlin} by an appropriate box-counting method that enters the
measurement of the fractal dimension. We study two genetic regulatory networks
({\it Saccharomyces cerevisiae} \cite{teichmann} and {\it Escherichai coli} \cite{zheng})
and show their self-similar and scale-free features, in extension to the datasets studied
by \cite{havlin}. Moreover, by a number of numerical results we support the conjecture that
self-similar scale-free networks are not assortative. From our simulations so far these
networks seem to be disassortative instead. We also find that the qualitative feature of
disassortativity is scale-invariant under renormalization, but it appears as an intrinsic feature
of the renormalization prescription, as even assortative networks become disassortative after
a sufficient number of renormalization steps.
\end{abstract}

\pacs{05.45.Df, 89.75.Hc, 87.16.Yc}
\maketitle

\section{\label{sec:intro}Introduction}

Until very recently, the celebrated properties of a scale-free
degree distribution seemed to be incompatible with self-similar
features of networks, in which the number of boxes $N_B$ of linear
size $\ell_B$ scales with $\ell_B$ according to a power-law
$N_B\sim  \ell_B^{-d_B}$, with an exponent that is given by the
fractal dimension $d_B$. Using the box-counting method, Song {\it {\it et al.}}
showed that many scale-free (SF) networks observed in nature can
have a fractal structure as well \cite{havlin}.
This result is striking, because the tiling and renormalization
according to the linear size of the boxes, in which all pairs of nodes
inside a box have mutual distance less than $\ell_B$, appear to
be physically relevant rather than being a formal procedure.
Therefore, the essential quantity in the tiling is the linear size
of the box, $\ell_B$, defined by $\ell_B-1$ being the maximal distance between
the nodes of the box.
In the renormalization procedure  the boxes are contracted to the
nodes of the renormalized network whose edges are the
interconnecting edges  between the boxes on the original network.

In this paper we study the genetic regulatory network of two well-known
organisms, {\it Saccharomyces cerevisiae} \cite{teichmann} and
{\it Escherichia coli} \cite{zheng}.
We first determine the degree-distribution $P(k)$, that is the probability for finding a node with degree $k$,
to read off the exponent $\gamma$ according to $P(k)\sim k^{-\gamma}$ \cite{Albert} in order to check that the networks
are scale-free. Next we measure $N_B/N$ for various box-sizes $\ell_B$ to obtain the fractal dimension
$d_B$ from $N_B/N\sim \ell_B^{-d_B}$. After renormalizing the networks according to the
procedure proposed in \cite{havlin}, we measure the scaling behavior of the degree $k^\prime$ according to
$k^\prime=s(\ell_B)k$, where $k^\prime$ stands for the degree of a node in the renormalized network, $k$
is the largest degree inside the box that was contracted to one node with degree $k^\prime$ in the renormalization
process, and $s(\ell_B)$ is assumed to scale like $s(\ell_B)\sim \ell_B^{-d_k}$ with a new exponent $d_k$.
The invariance of $\gamma$ under renormalization and the transformation behavior of the degree itself imply
the relation \cite{havlin}
\begin{equation}\label{1}
\gamma \;=\; 1\;+\;d_B/d_k
\end{equation}
between the exponents. Therefore we check this relation by measuring $d_B$, $d_k$ and comparing the
values of $\gamma$ from Eq.\ref{1} with the measured $\gamma$ from the degree distribution.

One of the important features of networks is their ``degree''  of
assortativity. The notion of assortative mixing was known from
epidemiology and ecology when it was introduced as a
characteristic feature of generic networks by Newman
\cite{newman}. Assortativity refers to correlations between
properties of adjacent nodes. One particular property is the
(in- or out-)degree of a node as the number of its (in- or
out-)going links, respectively. Degree-degree correlations can be
recorded as histograms; in order to facilitate the comparison
between networks of different size, they can be also characterized
by the Pearson coefficient. The Pearson coefficient is obtained
from the connected degree-degree correlation function
$\left<jk\right>-\left<j\right>\left<k\right>$ after normalizing
by its maximal value, which is achieved on a perfectly assortative
network. Here, $\left<jk\right>$ stands for the average of having
vertex degrees $j$ and $k$ at the end of an arbitrary edge. The
Pearson coefficient $r$ takes values between $-1\leq r \leq 1$, it
is positive for assortative networks ($r=1$ for complete
assortativity) and negative for disassortative ones. We have
measured this coefficient for a number of self-similar scale-free
networks and present the results below. The reason why this
feature is of interest in the present context is its relation to
the power-law or exponential behavior of $N_B(\ell_B)$. In
particular, we are interested in the question whether
disassortativity is scale-invariant on a qualitative level under
renormalization according to the prescription proposed in
\cite{havlin}, and why these properties go along. Disassortative
features in protein interaction networks were found and explained
by Maslov and Sneppen \cite{maslov} on the level of interacting
proteins and genetic regulatory interactions. According to  their
results links between highly connected nodes are systematically
suppressed, while those between highly-connected and low-connected
pairs of proteins are favored. In this way there is little
cross-talk between different functional modules of the cell and
protection against intentional attacks, since the failure of one
module is less likely to spread to another one. Also in
immunological networks one speaks of lock- and key-interactions
between molecular receptors and antigenic determinants
\cite{copelli}. In general, complementarity is essential for
pattern recognition interactions, underlying biological and
biochemical processes as well as for symbiotic species in
ecological networks. Of course, it is not at all obvious or necessary that
complementarity in ``internal'' (functional) properties should be
manifest in topological features like the degree-degree
correlations. Therefore we study the relation between
self-similarity and degree-assortativity in this paper.

\section{Measurements and Results}

For the genetic regulatory networks {\it Saccharomyces cerevisiae} \cite{teichmann} and
{\it Escherichia coli} \cite{zheng} we
observe a power-law
behavior of $N_B/N$ for $\ell_B > 3$ with $d_B =5.1\pm 0.3 $ for
{\it S. cerevisiae} and $d_B=3.4\pm0.2$, respectively. The
obtained degree-distributions are scale-free
and satisfy a power-law
 with exponent $\gamma \simeq 2.6$ for {\it S. cerevisiae}
and $\gamma\simeq 2.1$ for {\it E.coli}. The scaling relation (\ref{1}) between
the exponents $d_B$ and $d_k$ is
also satisfied within the error bars for both networks.

\begin{figure}
\includegraphics[width=8.5cm]{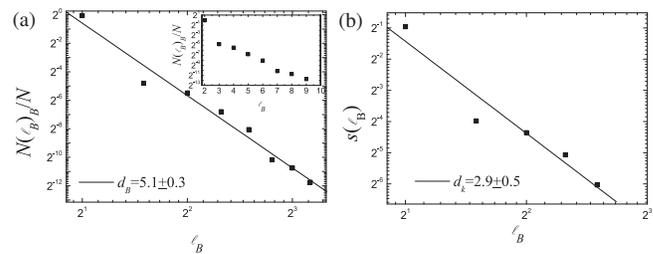}
\caption{\label{dBdk_YeastRegulatoryNet}  {\it
S.cerevisiae}: (a) Normalized number of boxes $N_B$ as a function
of linear box size $\ell_B$ to read off $d_B$ (b) Rescaling factor
$s(\ell_B)$ as function of the box size to read off $d_k$ }
\end{figure}

In Table \ref{tab:table1} we summarize the results also for some
additional networks, for which we list their properties of
self-similarity and disassortativity. If we confirm the property
of self-similarity it means not only the scaling behavior of
$N_B/N$ according to a power-law, but also the numerical
verification of the scaling relation of Eq.\ref{1} and the
invariance of $\gamma$ under renormalization\cite{havlin}. This is
more conclusive, because it is sometimes difficult to disentangle
exponential from power-law behavior of $N_B/N$ for networks with a
small diameter (for example see Fig.\ref{dBdk_YeastRegulatoryNet}
for the regulatory network of {\it S.cerevisiae} with an inset
that shows the same data points on a log-linear scale instead of
the log-log scale), whereas the scaling relation only holds for a
power-law of the decay, it is easier to prove or disprove.  A
confirmation of (dis)assortativity refers to histograms with
(negative) positive slope of next-neighbor degree-degree
correlations and/or a (negative) positive Pearson coefficient,
respectively. In most cases we measured the degree-degree
correlations also between nodes at distance $d=2,3$ as indicated
in the figures.

\begin{figure}
\includegraphics[width=8cm]{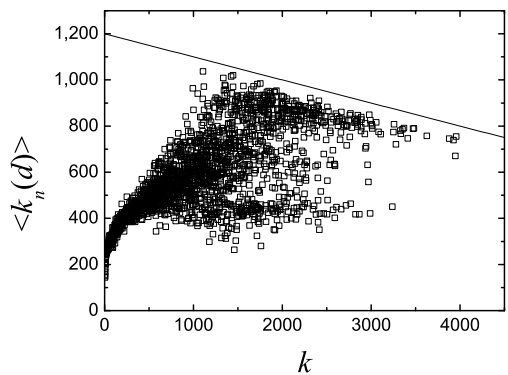}
\caption{\label{kknn_actor} Degree-degree correlation $\left<k_n(d)\right>$ for $d = 1$ against $k$ for the actor
network, showing initially assortative behavior for $k\preceq 1000$}
\end{figure}

The first four networks of Table \ref{tab:table1}
\begin{table}
\caption{\label{tab:table1}
Networks from datasets characterized by properties of self-similarity and disassortativity.}
\begin{ruledtabular}
\begin{tabular}{ccc}
Network&Self-similarity&\mbox{Disassortativity}\\
\hline
genetic reg.{\it S.cerevisia} \cite{teichmann}&yes Fig.\ref{dBdk_YeastRegulatoryNet}
&yes Fig.\ref{kknn_Yeast_Ecoli_RegulatoryNet} a\\
genetic reg.{\it E.coli} \cite{zheng}&yes Fig.\ref{dBdk_EcoliRegulatoryNet_Zheng_largedata}&yes
Fig.\ref{kknn_Yeast_Ecoli_RegulatoryNet} b\\
scient.collab. \cite{bara}& no Fig.\ref{ScientificCollaboration} a &no Fig.\ref{ScientificCollaboration} b\\
internet aut.sys.\cite{bara}& yes \cite{supply} &yes \cite{supply}\\
biochem.pathway{\it E.coli} \cite{bara}&yes \cite{havlin} &yes \cite{supply}\\
actor \cite{bara}&yes \cite{havlin} &(yes,$k>1000$)Fig.\ref{kknn_actor}\\
www \cite{bara}&yes \cite{havlin} &yes \cite{supply}\\
\end{tabular}
\end{ruledtabular}
\end{table}
refer to the genetic regulatory networks of
{\it S.cerevisia} and {\it E.coli}, the scientific collaboration network,
and the internet on the autonomous systems level. For these networks the properties of column 2 and 3
were examined
by us, while for the last three networks (the biochemical pathway network of {\it E.coli}, the actor network and the
world-wide-web), the self-similarity was
established before \cite{havlin}, and we studied their
property of disassortativity in addition. In particular the actor
network deserves some further comments. The actor-network is
self-similar \cite{havlin}, but its positive Pearson coefficient
suggests that it is assortative, in contrast to all other
self-similar networks we have studied so far. A closer look at its
next neighbor-degree-degree correlation (Fig.\ref{kknn_actor})
shows an assortative
behavior for degrees up to the order of 1000, but slowly decays
for larger degrees and becomes disassortative. The degree-degree
correlation between nodes at distance larger than 1 is decreasing with
degree $k$ for all $k$.

\begin{figure}
\includegraphics[width=8.5cm]{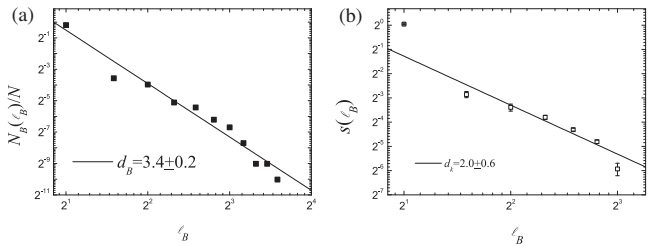}
\caption{\label{dBdk_EcoliRegulatoryNet_Zheng_largedata} Same as Fig.\ref{dBdk_YeastRegulatoryNet}, but for
{\it E.coli} to read off (a) $d_B$ (b) $d_k$}
\end{figure}

\begin{figure}
\includegraphics[width=8.5cm]{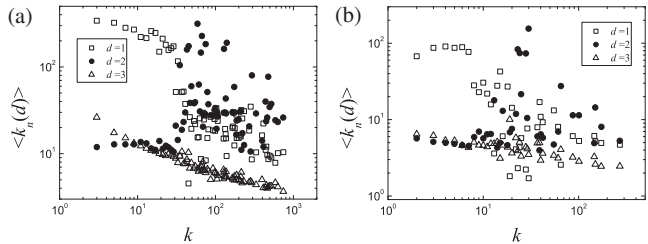}
\caption{\label{kknn_Yeast_Ecoli_RegulatoryNet} Degree-degree correlation $\left<k_n(d)\right>$ versus degree $k$
for distances $d =1,2,3$, (a) {\it S.cerevisiae}
and (b){\it E.coli}}
\end{figure}

Moreover, some comments are in order to the yeast-genetic
regulatory network with 3456 nodes and 14117 edges, (cf.
Fig.\ref{dBdk_YeastRegulatoryNet}). Since it has a diameter of 9,
the largest $\ell_B$ value for the tiling is 10. Therefore we have
only 8 data points available for the fit. Each point corresponds
to an average over 100 tiling configurations. Different tiling
configurations result from different starting seeds as well as the
random selection of neighbors during the tiling process. The data
point at $\ell_B = 3$ in Fig.\ref{dBdk_YeastRegulatoryNet} lies
clearly outside the fluctuations about the average over different
tiling configurations, thus outside the error bars, which are at
least two orders of magnitude smaller than the respective value of
$N_B/N$, so that they are not visible on the scale of the figure.
In Fig. \ref{dBdk_EcoliRegulatoryNet_Zheng_largedata}, we find
a similar behavior for {\it E.coli}. The deviation from the
power-law behavior at $\ell_B = 3$ goes along with an assortative
degree-degree correlation between nodes at distance $d=2$ as it is
seen from Fig.\ref{kknn_Yeast_Ecoli_RegulatoryNet}a and
Fig.\ref{kknn_Yeast_Ecoli_RegulatoryNet}b, showing the
degree-degree correlation of {\it S.cerevisiae} and {\it E.coli},
respectively, at distances $d =1,2,3$.
%
The data in Fig.\ref{kknn_Yeast_Ecoli_RegulatoryNet}
explicitly show the disassortative behavior at $d=1$ and $d=3$ for both {\it
S.cerevisiae} and {\it E.coli}. However, for $d=2$, we find that
there is a certain value of $k, k=k^*$, at which $\left<k_n\right>$
abruptly increases and slowly decreases for $k>k^*$. Here $k^*\sim
30$ for {\it S.cerevisiae} and $k^*\sim 10$ for {\it E.coli}.
These mixed properties of assortativity and disassortativity seem to go along with the deviation
from the power-law behavior of $N_B/N$. On a qualitative level, this is plausible if we focus on a hub
that should be present in a scale-free network. In an assortative network (assortative say at distance $d$,
for example $d=2$), this hub is likely connected to another hub within the distance $d$. If this hub is chosen
as a seed of a box in a tiling with linear box size $\ell_B > d$, we need much less boxes to cover the many
nodes in the neighborhood of the hub than in a disassortative network.

\begin{figure}
\includegraphics[width=8.5cm]{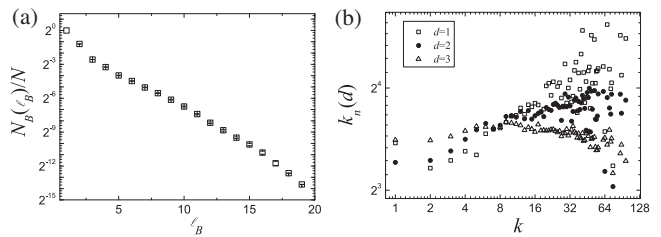}
\caption{\label{ScientificCollaboration} Scientific collaboration network (a) exponential decay of $N_B/N(\ell_B)$
(b) degree-degree correlation $\left<k_n(d)\right>$ for $d = 1,2,3$ against $k$}
\end{figure}

In a network which is assortative not only for a certain range of $k$, but for all $k$ and at distances
$d\geq 1$, like the scientific collaboration network, $N_B/N$ actually decays faster than power-like for
all $\ell_B$, as it is seen from the exponential fit of Fig.\ref{ScientificCollaboration}a.

\begin{figure}
\includegraphics[width=7cm]{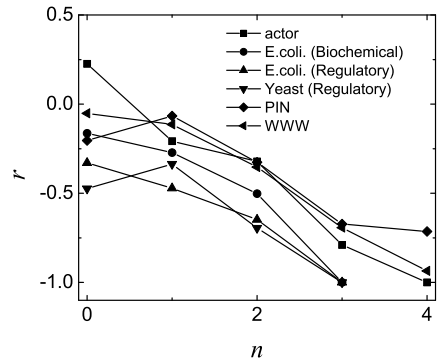}
\caption{\label{Pearson} Pearson coefficient for several data sets,
in particular the PIN-data taken from \cite{uetz}.}
\end{figure}

The scaling relation between the exponents $\gamma,
d_B$ and $d_k$ assumes the scale-invariance (under renormalization) of the degree
distribution, that is the invariance of the exponent $\gamma$. Similarly,
it is of interest how the disassortativity transforms under
renormalization (as defined in \cite{havlin}). As we see from Fig.\ref{Pearson},
\noindent
even networks like the scientific collaboration network (Fig.\ref{ScientificCollaboration}),
which are originally assortative, transform to more and more
disassortative ones under iterated renormalization.
(The number of renormalization steps is determined by the size of
the networks, in particular by its diameter. The final step is
achieved when the reduced network consists of just one node.)
Therefore the transformation behavior of disassortativity seems to
be an effect of the renormalization procedure rather than an
intrinsic self-similar property of the network. Similarly, we
measured the transformation behavior of the clustering coefficient
under renormalization of self-similar networks. As the data
\cite{supply} show, it is an invariant property of scale-free
networks, while it changes under renormalization for non-self-similar ones like the
Barab{\'a}si-Albert one \cite{BA}.

To summarize, we find numerical evidence that self-similar
scale-free networks are preferably disassortative in their
degree-degree correlations. For biological networks this result
may reflect the complementarity in interactions that is observed
on various levels.

\begin{acknowledgments}
One of us (H.M.-O.) would like to thank S.Havlin for drawing her attention
to self-similar networks at the COSIN-final meeting at Salou
(Spain) March 2005.
\end{acknowledgments}


\end{document}